%% file: main.tex
\documentclass[conference]{IEEEtran}
\IEEEoverridecommandlockouts

\usepackage{cite}
\usepackage{amsmath,amssymb,amsfonts}
\usepackage{algorithmic}
\usepackage{graphicx}
\usepackage{textcomp}
\usepackage{xcolor}
\usepackage{todonotes}
\usepackage{subfig}
\usepackage{soul}
\usepackage{amssymb}
\usepackage{booktabs}
\usepackage{multirow}
\usepackage[colorlinks=true,linkcolor=blue,citecolor=blue]{hyperref}
\def\BibTeX{{\rm B\kern-.05em{\sc i\kern-.025em b}\kern-.08em
    T\kern-.1667em\lower.7ex\hbox{E}\kern-.125emX}}
\begin{document}

\input{macro}

\title{Real-Time Low-Complexity Zak-OTFS\\with Superimposed Spread Pilots on GPUs
\vspace{-1mm}
\thanks{This work was supported in part by NSF under grants 2434131, 2443137, 2450567, 2503010, and 2554552, and an NVIDIA Academic Grant.
The work of Robert Calderbank, Danish Nisar, and Mudasir Sheikh is supported in part by NSF under grant 2342690 and by Cohere Technologies.}
}

\author{
Chung-Hsuan Tung,
Junyao Zheng,
Vineetha Yogesh$^*$\thanks{$^{*}$These authors contributed equally to this work.},
Mudasir Ahmad Sheikh$^*$,
Danish Nisar$^*$,\\
Robert Calderbank, and
Tingjun Chen\\
{Department of Electrical and Computer Engineering, Duke University, Durham, NC 27708, USA}\\
\vspace{-8mm}
}

\maketitle

\begin{abstract}
Orthogonal time frequency space (OTFS) modulation offers robust performance under high-mobility scenarios.
Spread pilots distribute pilot energy across the delay--Doppler grid and are superimposed with data symbols, enabling full-grid resource utilization.
However, pilot--data separation challenges the design of the real-time receiver pipeline: cross-ambiguity in channel sensing incurs high computational complexity, and pilot--data interference cancellation requires repeated processing through turbo iterations.
Leveraging the chirp structure of OTFS spread pilots, we reduce cross-ambiguity to a single 2D IFFT with element-wise chirp operations, achieving over $7000\times$ speedup compared to partially vectorized direct evaluation on GPUs at a frame size of $(\numDelayBin,\numDopplerBin)=(2053,31)$.
Our real-time GPU baseband processing pipeline for OTFS with superimposed spread pilots supports $(\numDelayBin,\numDopplerBin)=(2053,31)$ with one turbo iteration within a $1.03$\thinspace{\msec} latency budget.
We further show that larger frames increase processing gain, allowing lower pilot power and near-converged bit error rate (BER) after one turbo iteration.
\end{abstract}


\input{1_intro}
\input{2_sys}
\input{3_xamb}
\input{4_impl}

\input{5_eval}
\input{6_concl}

\bibliographystyle{IEEEtran}
\bibliography{reference}

\end{document}

%% file: macro.tex
%% COMMENTS
\definecolor{lightgray}{gray}{0.9}
\definecolor{lightblue}{rgb}{0.9,0.9,1}
\definecolor{LightMagenta}{rgb}{1,0.5,1}
\definecolor{red}{rgb}{1,0,0}
\definecolor{brightgreen}{rgb}{0.4, 1.0, 0.0}

\newcommand\couldremove[1]{{\color{lightgray} #1}}
\newcommand{\remove}[1]{}
\newcommand{\move}[2]{ {\textcolor{Purple}{ \bf --- MOVE #1: --- }} {\textcolor{Orchid}{#2}} }

%% FOR COMMENTS & EDITS
\newcommand{\hlc}[2][yellow]{ {\sethlcolor{#1} \hl{#2}} }
\newcommand\note[1]{\hlc[SkyBlue]{-- #1 --}} % highlighted notes of other colors.
% For colors info from xcolor package, check out:
% http://en.wikibooks.org/wiki/LaTeX/Colors

\newcommand\mynote[1]{\hlc[yellow]{#1}}
\newcommand\tingjun[1]{\hlc[yellow]{TC: #1}}
\newcommand\andy[1]{\hlc[pink]{Andy: #1}}
\newcommand\bill[1]{\hlc[cyan!50]{Bill: #1}}
\newcommand\change[1]{{\color{blue} {#1}}}

% Revision highlights
\newcommand{\TODO}[1]{\textcolor{red}{#1}}
\newcommand{\revise}[1]{\textcolor{blue}{#1}}

\newcommand{\myparatight}[1]{\vspace{0.5ex}\noindent\textbf{#1~}}

\newcommand{\cmark}{\ding{51}}%
\newcommand{\xmark}{\ding{55}}%
\newcommand{\greencheck}{\color[HTML]{3C8031}{\cmark}}
\newcommand{\redcross}{\color[HTML]{ED1B23}{\xmark}}
\newcommand{\greenno}{\color[HTML]{3C8031}{\textbf{No}}}
\newcommand{\redyes}{\color[HTML]{ED1B23}{\textbf{Yes}}}
\newcommand{\greenlow}{\color[HTML]{3C8031}{\textbf{Low}}}
\newcommand{\redhigh}{\color[HTML]{ED1B23}{\textbf{High}}}

\newcommand*\circledwhite[1]{\tikz[baseline=(char.base)]{
            \node[shape=circle,draw,inner sep=0.6pt] (char) {\scriptsize{#1}};}}

\newcommand*\circled[1]{\tikz[baseline=(char.base)]{
            \node[shape=circle,draw,fill=black,text=white,inner sep=1pt] (char) {\scriptsize{#1}};}}

%% paper-specific variable/notation
% \newcommand{\name}{{SS-CGA}}
% \newcommand{\namebf}{{\textbf{SS-CGA}}}
% \newcommand{\namesc}{{\sf Savannah-sc}}
% \newcommand{\namescbf}{{\sf\textbf{Savannah-sc}}}
% \newcommand{\namemc}{{\sf Savannah-mc}}
% \newcommand{\namemcbf}{{\sf\textbf{Savannah-mc}}}
% \newcommand{\namemcidx}[1]{{\sf Savannah-{#1}c}}
% \newcommand{\namemcidxbf}[1]{{\sf\textbf{Savannah-{#1}c}}}

% \newcommand{\agora}{Agora}
% \newcommand{\agorabf}{\textbf{Agora}}

% \newcommand{\armavec}{\sf Savannah-mc (arma-vec)}
% \newcommand{\armacube}{\sf Savannah-mc (arma-cube)}

%%This is a special cell that allows using \\ inside the cell to add a new line
\newcommand{\specialcell}[2][c]{%
  \begin{tabular}[#1]{@{}c@{}}#2\end{tabular}}

\newcommand{\iu}{{j}}

\newcommand{\littlesum}{\mathop{\textstyle\sum}}
\newcommand{\littleint}{\mathop{\textstyle\int}}

% config
\newcommand{\siso}{SISO}
\newcommand{\mimoTwoByTwo}{2$\times$2}
\newcommand{\mimoFourByFour}{4$\times$4}

% lib
\newcommand{\bbdev}{\textsf{bbdev}}

% PHY
\newcommand{\scs}{{\Delta f}}
\newcommand{\scNum}{N_{\textrm{sc}}}
\newcommand{\sampRate}{F_{\textrm{s}}}
\newcommand{\fftSize}{N_{\textrm{FFT}}}
\newcommand{\chMat}{{\textbf{H}}}
\newcommand{\chVec}{{\textbf{h}}}
\newcommand{\precodeMat}{{\textbf{P}}}

% units
\newcommand{\usec}{$\mu$s} % us or usec
\newcommand{\msec}{ms}     % ms or msec

% OTFS parameters
\newcommand{\numDelayBin}{M}
\newcommand{\numDopplerBin}{N}
\newcommand{\delay}{\tau}
\newcommand{\DopplerShift}{\nu}
\newcommand{\delayPeriod}{\delay_{p}}
\newcommand{\DopplerPeriod}{\DopplerShift_{p}}
\newcommand{\delayRes}{\Delta\delay} % delay resolution
\newcommand{\DopplerRes}{\Delta\DopplerShift} % Doppler resolution
\newcommand{\freqSpacing}{\Delta f} % equivalent to OFDM subcarrier spacing, = Doppler period
\newcommand{\pathIdx}{i}
\newcommand{\pathDelay}{\tau_{\pathIdx}}
\newcommand{\pathDoppler}{\nu_{\pathIdx}}
\newcommand{\delayIdx}{k}
\newcommand{\DopplerIdx}{l}
\newcommand{\pilotDelayIdx}{K_0}
\newcommand{\pilotDopplerIdx}{L_0}
\newcommand{\pathDelayIdx}{{\delayIdx}_{\pathIdx}}
\newcommand{\pathDopplerIdx}{{\DopplerIdx}_{\pathIdx}}

\newcommand{\chEleDD}{H_{\textrm{dd}}}
\newcommand{\chMatDD}{\textbf{H}_{\textrm{dd}}}
\newcommand{\chEleDDEst}{\widehat{H}_{\textrm{dd}}}
\newcommand{\chMatDDEst}{\widehat{\textbf{H}}_{\textrm{dd}}}
\newcommand{\chMatDDEstHerm}{\widehat{\textbf{H}}^\textrm{H}_{\textrm{dd}}}

% OTFS data structure parameters
\newcommand{\chMatDDRowIdx}{q}
\newcommand{\chMatDDColIdx}{r}
\newcommand{\chMatDDColIdxMapped}[1]{{\chMatDDColIdx}_{\pathIdx}(#1)}
\newcommand{\chMatDDColIdxMappedRowIdx}{\chMatDDColIdxMapped{\chMatDDRowIdx}}
\newcommand{\chMatDDRowIdxMapped}[1]{{\chMatDDRowIdx}_{\pathIdx}(#1)}
\newcommand{\chMatDDRowIdxMappedColIdx}{\chMatDDRowIdxMapped{\chMatDDColIdx}}
\newcommand{\sparseCoe}{D}
\newcommand{\mvmCoe}[2]{\sparseCoe_{{#1},{#2}}}
\newcommand{\coePathRowIdx}{\mvmCoe{\pathIdx}{\chMatDDRowIdx}}
\newcommand{\chMatDDColDelayBinOffset}{{d}_{\delayIdx}(\pathIdx)}
\newcommand{\chMatDDColDopplerBinOffset}{{d}_{\DopplerIdx}(\pathIdx)}
\newcommand{\chMatDDRowDelayBinOffset}{{\delta}_{\delayIdx}(\pathIdx)}
\newcommand{\chMatDDRowDopplerBinOffset}{{\delta}_{\DopplerIdx}(\pathIdx)}
\newcommand{\chPathGain}{h_{\pathIdx}}
\newcommand{\phCompFactor}{\gamma_{\pathIdx,{\chMatDDRowIdx}}}
\newcommand{\phTwisted}{\phi_{\pathIdx,{\chMatDDRowIdx}}}
\newcommand{\rawDelayOffset}{A_{\pathIdx,{\chMatDDRowIdx}}}
\newcommand{\DelayOffsetWrapCnt}{n_{\pathIdx,{\chMatDDRowIdx}}}

\newcommand{\chNoiseDDVec}{\textbf{n}_{\textrm{dd}}}
\newcommand{\chNoiseDDMat}{\textbf{N}_{\textrm{dd}}}
\newcommand{\chNoiseDDMatEle}{n_{\textrm{dd}}}
\newcommand{\chNoiseTd}{n}

\newcommand{\chEleMMSE}{H_{\textrm{LMMSE}}}
\newcommand{\chMatMMSE}{\textbf{H}_{\textrm{LMMSE}}}

\newcommand{\chPhy}{h_{\textrm{phy}}}
\newcommand{\chEff}{h_{\textrm{eff}}}
\newcommand{\chEffEstMat}{\widehat{\textbf{h}}_{\textrm{eff}}}
\newcommand{\chEffEstMatEle}{\widehat{h}_{\textrm{eff}}}
\newcommand{\chEffMat}{\mathbf{h}_{\textrm{eff}}}
\newcommand{\chEffMatEle}{h_{\textrm{eff}}}

\newcommand{\rxFilter}{w_\mathrm{rx}}
\newcommand{\txFilter}{w_\mathrm{tx}}

\newcommand{\sigTx}{x}
\newcommand{\sigTxSp}{\sigTx_\mathrm{sp}}
\newcommand{\sigTxDt}{\sigTx_\mathrm{dt}}
\newcommand{\sigTxDD}{x_{\textrm{dd}}}
\newcommand{\sigTxDDVec}{\mathbf{x}_{\textrm{dd}}}
\newcommand{\sigTxDDVecEle}{x_{\textrm{dd}}}
\newcommand{\sigTxDDVecEst}{\widehat{\mathbf{x}}_{\textrm{dd}}}
\newcommand{\sigTxDDMatEle}{x_{\textrm{dd}}}
\newcommand{\sigTxDDMat}{\mathbf{X}_{\textrm{dd}}}
\newcommand{\sigTxDDMatEst}{\widehat{\mathbf{X}}_{\textrm{dd}}}

\newcommand{\sigTxDDPtMat}{\mathbf{X}_{\textrm{pt,dd}}}
\newcommand{\sigTxDDPtMatEle}{x_{\textrm{pt,dd}}}
\newcommand{\sigTxDDSpMat}{\mathbf{X}_{\textrm{sp,dd}}}
\newcommand{\sigTxDDSpMatEle}{x_{\textrm{sp,dd}}}
\newcommand{\sigTxDDSpVec}{\mathbf{x}_{\textrm{sp,dd}}}
\newcommand{\sigTxDDSpVecEst}{\widehat{\mathbf{x}}_{\textrm{sp,dd}}}
\newcommand{\sigTxDDDtMat}{\mathbf{X}_{\textrm{dt,dd}}}
\newcommand{\sigTxDDDtMatEle}{x_{\textrm{dt,dd}}}
\newcommand{\sigTxDDDtMatEst}{\widehat{\mathbf{X}}_{\textrm{dt,dd}}}
\newcommand{\sigTxDDDtVec}{\mathbf{x}_{\textrm{dt,dd}}}
\newcommand{\sigTxDDDtVecEst}{\widehat{\mathbf{x}}_{\textrm{dt,dd}}}
\newcommand{\sigTxDDDtVecEleEst}{\widehat{x}_{\textrm{dt,dd}}}
\newcommand{\sigTxDDDtMatEstDemod}{\breve{\mathbf{X}}_{\textrm{dt,dd}}}
\newcommand{\sigTxDDDtVecEstDemod}{\breve{\mathbf{x}}_{\textrm{dt,dd}}}
\newcommand{\sigTxDDDtVecEstDemodEle}{\breve{x}_{\textrm{dt,dd}}}

\newcommand{\sigTxTd}{s}
\newcommand{\sigTxTdVecEle}{s}
\newcommand{\sigTxTdVec}{\mathbf{s}}
\newcommand{\sigTxTdMatEle}{s}
\newcommand{\sigTxTdMat}{\mathbf{S}}

\newcommand{\sigRxDD}{y_{\textrm{dd}}}
\newcommand{\sigRxDDVec}{\mathbf{y}_{\textrm{dd}}}
\newcommand{\sigRxDDMatEle}{y_{\textrm{dd}}}
\newcommand{\sigRxDDMat}{\mathbf{Y}_{\textrm{dd}}}
\newcommand{\sigRxTd}{r}
\newcommand{\sigRxTdVec}{\mathbf{y}}
\newcommand{\sigRxTdMat}{\mathbf{Y}}

\newcommand{\sigRxDDSpMat}{\mathbf{Y}_{\textrm{sp,dd}}}
\newcommand{\sigRxDDSpMatEst}{\widehat{\mathbf{Y}}_{\textrm{sp,dd}}}
\newcommand{\sigRxDDSpMatEle}{y_{\textrm{sp,dd}}}
\newcommand{\sigRxDDSpVec}{\mathbf{y}_{\textrm{sp,dd}}}
\newcommand{\sigRxDDSpVecEst}{\widehat{\mathbf{y}}_{\textrm{sp,dd}}}
\newcommand{\sigRxDDDtMat}{\mathbf{Y}_{\textrm{dt,dd}}}
\newcommand{\sigRxDDDtMatEst}{\widehat{\mathbf{Y}}_{\textrm{dt,dd}}}
\newcommand{\sigRxDDDtMatEle}{y_{\textrm{dt,dd}}}
\newcommand{\sigRxDDDtVec}{\mathbf{y}_{\textrm{dt,dd}}}
\newcommand{\sigRxDDDtVecEst}{\widehat{\mathbf{y}}_{\textrm{dt,dd}}}

\newcommand{\bitTx}{\mathbf{b}}
\newcommand{\bitRx}{\widehat{\mathbf{b}}}

\newcommand{\chMaxDopplerFreq}{\nu_{\text{max}}}

\newcommand{\bandwidth}{B}
\newcommand{\frameTime}{T}
\newcommand{\numPath}{P} % corresponds to the number of non-zero entries in the channel matrix
\newcommand{\coherenceTime}{T_c}

\newcommand{\heffThres}{\theta}
\newcommand{\chPathThresSet}{\widehat{\Omega}}

% OTFS CGA parameters
\newcommand{\noiseCovMat}{\mathbf{R}_{n}}
\newcommand{\cgaIter}{\xi}
\newcommand{\cgaIters}{\Xi}
\newcommand{\sqResNorm}{c_\textrm{norm}}

% OTFS exponential terms
\newcommand{\kernelFftMatEle}{E_{\text{Zak}}}
\newcommand{\kernelFftMat}{\mathbf{E}_{\text{Zak}}}
\newcommand{\twistMatEle}{E_{\text{twist}}}
\newcommand{\twistMat}{\mathbf{E}_{\text{twist}}}

% OTFS indexing
\newcommand{\delayIdxSet}{\mathcal{\numDelayBin}}
\newcommand{\DopplerIdxSet}{\mathcal{\numDopplerBin}}
\newcommand{\pathIdxSet}{\mathcal{\numPath}}
\newcommand{\gridFlatIdxSet}{\mathcal{Q}} % index set for flattened grid (0, 1, ..., mn-1)

% Other helpful
\newcommand{\jimg}{j}
\newcommand{\defn}{\triangleq}
\newcommand{\mymod}[2]{\langle #1 \rangle_{#2}}

\newcommand{\myAbs}[1]{\left|{#1}\right|}
\newcommand{\myAng}[1]{\angle{#1}}
\newcommand{\myConjugate}[1]{{#1}^{*}}
\newcommand{\myTranspose}[1]{{#1}^{\top}}
\newcommand{\myHermitian}[1]{{#1}^{H}}
\newcommand{\myIsFunc}[1]{\mathbf{1}\{#1\}}

\newcommand{\bw}{B}
\newcommand{\carrierFreq}{f_{c}}
\newcommand{\carrierWave}{\lambda}

\newcommand{\csiMat}{\mathbf{H}}

\newcommand{\SNR}{\mathsf{SNR}}

\newcommand{\bigO}{\mathcal{O}} % Open at top left

\newcommand{\pilotEnergy}{E_\mathrm{p}}
\newcommand{\dataEnergy}{E_\mathrm{d}}
\newcommand{\pdr}{\rho}

\newcommand{\chirpSlope}{q}
\newcommand{\chirpFilter}{w}
\newcommand{\twistConv}{*_{\sigma}}
\newcommand{\twistConvPeriodic}{{\circledast}_{\sigma}}
\newcommand{\xamb}[2]{A_{#1,#2}}
\newcommand{\turboIter}{u}
\newcommand{\numTurboIter}{U}
\newcommand{\atTurboIter}[1]{#1^{(\turboIter)}}
\newcommand{\atLastTurboIter}[1]{#1^{(\turboIter-1)}}

%% ---------------------------------------------------------------------------
%% Macros for chirp-structure-based FFT XAmb
%% ---------------------------------------------------------------------------
\newcommand{\MN}{\numDelayBin\numDopplerBin}
\newcommand{\omegaMN}{\omega_{\MN}}                  % e^{j2pi/(MN)}
\newcommand{\omegaM}{\omega_{\numDelayBin}}          % e^{j2pi/M}
\newcommand{\omegaN}{\omega_{\numDopplerBin}}        % e^{j2pi/N}
\newcommand{\modinv}[2]{\langle #1\rangle_{#2}^{-1}} % inverse of #1 modulo #2
\newcommand{\gaussSum}{\mathcal{G}}                  % quadratic Gauss sum
\newcommand{\delayIdxPilot}{\delayIdx_{\mathrm{P}}}  % point-pilot delay index
\newcommand{\DopplerIdxPilot}{\DopplerIdx_{\mathrm{P}}} % point-pilot Doppler index
\newcommand{\delayOff}{\bar{\delayIdx}}              % k - k_p
\newcommand{\DopplerOff}{\bar{\DopplerIdx}}          % l - l_p
\newcommand{\pilotConst}{\kappa}                     % pilot global constant
\newcommand{\pilotPhase}{\Phi}                       % pilot phase polynomial
\newcommand{\pc}[1]{c_{#1}}                          % coefficient of P
\newcommand{\gM}{\gamma_{\numDelayBin}}
\newcommand{\gN}{\gamma_{\numDopplerBin}}
\newcommand{\outPhase}{\Theta}                       % output chirp phase
\newcommand{\sigRxDechirp}{\tilde{Y}}                % de-chirped frame
\newcommand{\idxM}{a}                                % delay-bin index map
\newcommand{\idxN}{b}                                % Doppler-bin index map

%% ---------------------------------------------------------------------------
%% Macros for SVD-based Corr XAmb (replaces the previous block)
%% ---------------------------------------------------------------------------
\newcommand{\rank}{\Gamma}                                 % rank of twist matrix
\newcommand{\twistCompIdx}{\gamma}                         % SVD component index
\newcommand{\delayIdxSp}{\tilde{\delayIdx}}                % pilot delay index
\newcommand{\DopplerIdxSp}{\tilde{\DopplerIdx}}            % pilot Doppler index
\newcommand{\delayIdxSetCtr}{\mathcal{\numDelayBin}_{\mathrm{c}}}   % centered delay set
\newcommand{\DopplerIdxSetCtr}{\mathcal{\numDopplerBin}_{\mathrm{c}}} % centered Doppler set
\newcommand{\twistFacLMat}{\mathbf{G}}                     % left factor (scaled left sing. vec.)
\newcommand{\twistFacLVec}[1]{\mathbf{g}_{#1}}
\newcommand{\twistFacLVecEle}[1]{g_{#1}}
\newcommand{\twistFacRMat}{\mathbf{V}}                     % right singular vectors
\newcommand{\twistFacRVec}[1]{\mathbf{v}_{#1}}
\newcommand{\twistFacRVecEle}[1]{v_{#1}}
\newcommand{\sigTxDDSpCalMat}[1]{\mathbf{X}_{\textrm{sp,dd},#1}}  % calibrated pilot
\newcommand{\sigTxDDSpCalMatEle}[1]{x_{\textrm{sp,dd},#1}}
\newcommand{\corr}[2]{\mathcal{C}_{#1,#2}}                 % 2D cross-correlation

% Convenience macros
\newcommand{\dd}{\textrm{dd}}
\newcommand{\td}{\textrm{td}}
\newcommand{\phy}{\textrm{phy}}
\newcommand{\eff}{\textrm{eff}}
\newcommand{\defeq}{\overset{\Delta}{=}}

\newcommand{\fft}{\mathcal{F}_\textrm{2D}}
\newcommand{\ifft}{\fft^{-1}}

%% file: 1_intro.tex
\vspace{-1mm}
\section{Introduction}

Orthogonal time frequency space (OTFS) modulation, which embraces channel Doppler effects in its signal representation, is a promising waveform for high-mobility scenarios in next-generation (NextG) wireless networks.
OTFS modulates information symbols on the delay--Doppler (DD)-domain grid, which can incur higher receiver complexity than conventional orthogonal frequency-division multiplexing (OFDM), due to large-scale matrix operations for channel estimation and equalization.
Zak-OTFS~\cite{mohammed2023otfs} maps DD-domain symbols to time-domain waveforms through the inverse Zak transform.

In Zak-OTFS systems, a point pilot protected by a guard region allows direct read-off of the DD-domain effective channel taps, but also occupies DD resources that could otherwise carry data.
Spread pilots, instead, distribute the pilot across the entire DD grid, enabling full-grid utilization when superimposed with data symbols~\cite{ubadah2024zak}.
Channel read-off then requires computing the cross-ambiguity between the received signal and the spread pilot with a direct complexity of $\bigO(\numDelayBin^2\numDopplerBin^2)$ for a DD-domain grid with dimension $(\numDelayBin, \numDopplerBin)$, while the pilot and data share the available transmit power.
Canceling the pilot--data interference further requires turbo iterations~\cite{jayachandran2024zak}, which repeatedly invoke the receiver digital signal processing (DSP) stages and increases processing latency.

General-purpose computing platforms are increasingly adopted in wireless systems for baseband processing, trading compute efficiency for flexibility.
GPU acceleration offers a promising approach to balancing programmability and processing performance.
Recent studies achieved real-time baseband processing of OFDM signals on CPUs~\cite{qi2024savannah} and point-pilot-based OTFS signals on GPUs~\cite{zheng2026real}; the latter exploits DD-domain channel sparsity and GPU parallelism.
However, this sparsity is revealed only after channel sensing and thus cannot reduce the cost of cross-ambiguity for spread pilots.

This work exploits the structure of spread pilots, rather than the channel sparsity, for low-complexity channel sensing.
We show that its quasi-periodic cross-ambiguity reduces exactly to a chirp-modulated 2D inverse discrete Fourier transform (IDFT) and integrate the resulting acceleration into a GPU-accelerated, real-time Zak-OTFS receiver.

In summary, this paper makes the following contributions:
\begin{itemize}
    \item We exploit the spread-pilot chirp structure to reduce cross-ambiguity in Zak-OTFS to a 2D IDFT (CS-FFT), lowering the complexity from $\bigO(\numDelayBin^2\numDopplerBin^2)$ to $\bigO(\numDelayBin\numDopplerBin\log\numDelayBin\numDopplerBin)$.
    We also develop an FFT-based formulation using singular value decomposition (SVD) of the phase term (SVD-FFT) in $\bigO(\min(\numDelayBin,\numDopplerBin)\numDelayBin\numDopplerBin\log\numDelayBin\numDopplerBin)$.
    \item We implement an end-to-end, GPU-accelerated Zak-OTFS receiver with superimposed spread pilots and turbo pilot--data interference cancellation.
    Using CS-FFT cross-ambiguity, the receiver achieves real-time processing within the frame duration $\frameTime=1.03$\thinspace{\msec} for frame sizes up to $(\numDelayBin,\numDopplerBin)=(2053,31)$ with one turbo iteration.
    \item We characterize the reliability trade-off under a latency constraint and show that, as the number of delay bins $\numDelayBin$ increases, the bit error rate (BER) converges rapidly with the first turbo iteration capturing most of the achievable gain, while the BER-minimizing pilot power decreases.
\end{itemize}

%% file: 2_sys.tex
\section{System Model: Zak-OTFS with Spread Pilots}
\label{sec:sys}

\begin{figure}[!t]
    \centering
    \includegraphics[width=\columnwidth]{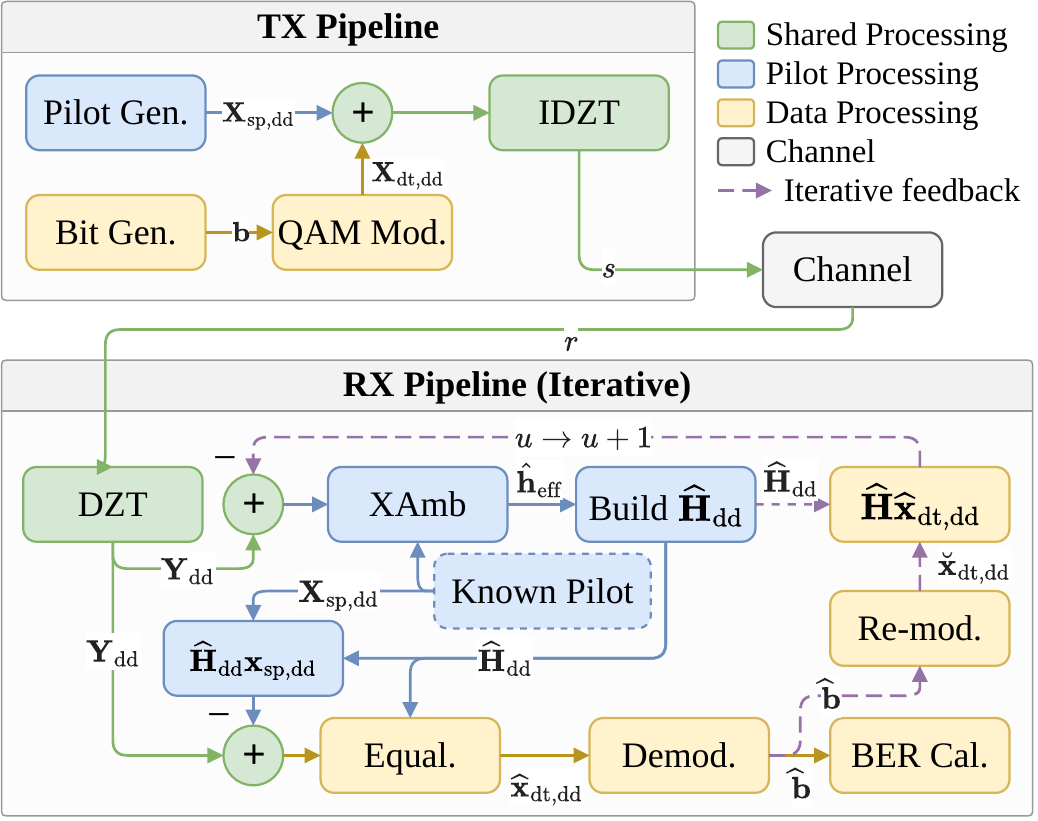}
    \caption{OTFS processing pipeline with superimposed spread pilots and turbo iterations.
    At the initial pass ($\turboIter=0$), $\sigTxDDDtVecEstDemod=\mathbf{0}$; the BER is computed against the transmitted bits $\bitTx$.
    Power scaling is omitted for clarity; see \eqref{eq:superimpose}, \eqref{eq:pilot_cancel}, \eqref{eq:turbo_data_cancel}.
    XAmb: Cross-ambiguity; Mod/Demod: Modulation/demodulation; Equal: Equalization.
    }
    \label{fig:pipeline}
    \vspace{-1.5mm}
\end{figure}

The carrier waveform of Zak-OTFS modulation, i.e., a delay--Doppler (DD) pulse, is a localized, quasi-periodic function specified by the delay period $\delayPeriod$ and the Doppler period $\DopplerPeriod = 1/\delayPeriod$, which together define the period lattice
$\Lambda_p = \{(n\delayPeriod, m\DopplerPeriod) \,|\, n,m \in \mathbb{Z}\}$.
The pulse is specified over the fundamental region
$\mathcal{D}_0
=
\{
    (\delay,\DopplerShift)
    \,|\,
    0 \le \delay < \delayPeriod,
    \ 
    0 \le \DopplerShift < \DopplerPeriod
\}$
and extends quasi-periodically over the DD plane as
\begin{equation}\label{eq:qp_ext_cont}
x_{\dd}(\delay + n\delayPeriod, \DopplerShift + m\DopplerPeriod)
=
e^{\jimg2\pi n \DopplerShift \delayPeriod}\, x_{\dd}(\delay,\DopplerShift),
\ 
\forall n,m \in \mathbb{Z}.
\end{equation}
A DD function has a time-domain (TD) counterpart only if it satisfies \eqref{eq:qp_ext_cont}.
In the TD, the DD pulse becomes a train of pulses multiplied by a complex tone, hence called a \emph{pulsone}~\cite{ubadah2024zak}.

Fig.~\ref{fig:pipeline} shows the considered transmitter and receiver pipeline with superimposed spread pilots~\cite{ubadah2024zak} and turbo iterations~\cite{jayachandran2024zak}.

\subsection{Zak-OTFS Modulation}\label{subsec:sys_mod}
A Zak-OTFS frame spans $\numDelayBin$ delay bins and $\numDopplerBin$ Doppler bins, indexed by
$\delayIdx\in\delayIdxSet=\{0,\ldots,\numDelayBin-1\}$ and
$\DopplerIdx\in\DopplerIdxSet=\{0,\ldots,\numDopplerBin-1\}$, respectively.
The corresponding TD frame occupies duration
$\frameTime=\numDopplerBin\delayPeriod$ and bandwidth
$\bandwidth=\numDelayBin\DopplerPeriod$, carrying
$\numDelayBin\numDopplerBin=\bandwidth\frameTime$ symbols.
We superimpose the spread pilot $\sigTxSp[\delayIdx,\DopplerIdx]$ and data symbols $\sigTxDt[\delayIdx,\DopplerIdx]$ as
\begin{equation}\label{eq:superimpose}
\sigTx[\delayIdx,\DopplerIdx]
=
\sqrt{\pilotEnergy}\,\sigTxSp[\delayIdx,\DopplerIdx]
+
\sqrt{\dataEnergy}\,\sigTxDt[\delayIdx,\DopplerIdx],
\end{equation}
where $\pilotEnergy$ and $\dataEnergy$ are the average pilot and data powers,
respectively, with pilot-to-data power ratio (PDR)
$\pdr\defn\pilotEnergy/\dataEnergy$~\cite{ubadah2024zak}.
The data symbols are drawn from a unit-energy constellation $\mathcal{S}$.
Without loss of generality, we set $\dataEnergy=1$, such that
$\pilotEnergy=\pdr$.
The symbols are placed into a discrete quasi-periodic DD signal
$x_{\dd}[\cdot,\cdot]$ by
\begin{equation}\label{eq:qp_ext_disc}
\sigTxDD[\delayIdx + n\numDelayBin, \DopplerIdx + m\numDopplerBin]
\defn
e^{\jimg2\pi \frac{n\DopplerIdx}{N}}\, \sigTx[\delayIdx,\DopplerIdx],
\quad
n,m \in \mathbb{Z}.
\end{equation}

A point pilot over the fundamental DD grid is defined as
$
x_{\textrm{pt}}[\delayIdx,\DopplerIdx]
=
\delta[\delayIdx-\delayIdxPilot]\delta[\DopplerIdx-\DopplerIdxPilot]
$,
where $(\delayIdxPilot,\DopplerIdxPilot)$ denotes the pilot location; its quasi-periodic extension $\sigTxDDPtMatEle[\delayIdx,\DopplerIdx]$ follows \eqref{eq:qp_ext_disc} and is $\numDelayBin\numDopplerBin$-periodic in both $(\delayIdx,\DopplerIdx)$.
The spread pilot is obtained by
\begin{equation}\label{eq:spread_pilot}
\sigTxDDSpMatEle[\delayIdx,\DopplerIdx]
\defn
\chirpFilter[\delayIdx,\DopplerIdx]
\ 
\twistConvPeriodic
\ 
\sigTxDDPtMatEle[\delayIdx,\DopplerIdx],
\ 
\forall \delayIdx,\DopplerIdx \in \mathbb{Z},
\end{equation}
where $\twistConvPeriodic$ denotes $MN$-periodic twisted
convolution~\cite{ubadah2024zak}, and
\begin{equation}\label{eq:chirp_filter}\textstyle
\chirpFilter[\delayIdx,\DopplerIdx]
=
\frac{1}{\sqrt{\numDelayBin\numDopplerBin}}
\cdot
e^{\jimg2\pi\frac{\chirpSlope(\delayIdx^2+\DopplerIdx^2)}{\numDelayBin\numDopplerBin}},
\ 
\forall \delayIdx,\DopplerIdx \in \mathbb{Z},
\end{equation}
is an $MN$-periodic discrete chirp filter with slope $\chirpSlope\in\mathbb{Z}$.
The spread pilot symbols $x_{\mathrm{sp}}[\delayIdx,\DopplerIdx]$ are given by $\sigTxDDSpMatEle[\delayIdx,\DopplerIdx]$ over the fundamental DD grid.

The discrete signal $\sigTxDD[\delayIdx,\DopplerIdx]$ is lifted onto the lattice
$
\Lambda_{\dd}
=
\{
    (\delayIdx\delayPeriod/\numDelayBin,\, \DopplerIdx\DopplerPeriod/\numDopplerBin)
    \,|\,
    \delayIdx,\DopplerIdx \in \mathbb{Z}
\}
$ to form a continuous DD signal
\begin{equation}\label{eq:lift}\textstyle
\sigTxDD(\delay,\DopplerShift)
\defn
\sum_{\delayIdx,\DopplerIdx \in \mathbb{Z}}
    \sigTxDD[\delayIdx,\DopplerIdx]\,
    \delta\!\left(\delay - \frac{\delayIdx\delayPeriod}{\numDelayBin}\right)
    \delta\!\left(\DopplerShift - \frac{\DopplerIdx\DopplerPeriod}{\numDopplerBin}\right),
\end{equation}
where $\delta(\cdot)$ denotes the  Dirac delta.
The lifted DD signal is pulse-shaped by the transmit pulse $\txFilter(\tau,\nu)$ through twisted convolution as
$
\sigTxDD^{\txFilter}(\delay,\DopplerShift)
\defn
\left(\txFilter\twistConv\sigTxDD\right)(\delay,\DopplerShift),
$
where $\twistConv$ denotes twisted convolution.\footnote{$
(a \twistConv b)(\tau,\nu)
\defn
\iint
    a(\tau',\nu')\,
    b(\tau-\tau', \nu-\nu')\,
    e^{\jimg2\pi \nu'(\tau-\tau')}\,
    d\tau'\,
    d\nu'
$
for two DD functions $a(\tau,\nu)$ and $b(\tau,\nu)$.}
The transmitted TD waveform is then obtained by the inverse Zak transform as
$
\sigTxTd(t)
=
\mathcal{Z}_t^{-1}
\left(\sigTxDD^{\txFilter}(\delay,\DopplerShift)\right).
$
In discrete time, ${\sigTxTdVecEle}[i]$ is generated by the inverse discrete Zak transform (IDZT)~\cite{lampel2022otfs}:
\begin{equation}\label{eq:idzt}\textstyle
{\sigTxTdVecEle}[i]
=
\frac{1}{\sqrt{\numDopplerBin}}
\sum_{\DopplerIdx=0}^{\numDopplerBin-1}
    {\sigTxDDMatEle}[\mymod{i}{\numDelayBin},{\DopplerIdx}]
    \cdot
    e^{\jimg 2\pi \frac{\left\lfloor i/\numDelayBin \right\rfloor}{\numDopplerBin}\DopplerIdx},
\end{equation}
where $\mymod{\cdot}{\numDelayBin}$ denotes the modulo $\numDelayBin$ operation and $i \in \{0,1,\ldots,{\numDelayBin}{\numDopplerBin}-1\}$.

\subsection{Zak-OTFS Receiver}\label{subsec:sys_rx}

Propagating $\sigTxTd(t)$ through the physical channel with DD spreading function $\chPhy(\delay,\DopplerShift)$ yields the received TD waveform
\begin{equation}\label{eq:td_channel}\textstyle
\sigRxTd(t)
=
\iint
    \chPhy(\delay,\DopplerShift)\,
    \sigTxTd(t-\delay)\,
    e^{\jimg2\pi \DopplerShift (t-\delay)}\,
    d\delay\,
    d\DopplerShift
+
\chNoiseTd(t),
\end{equation}
where $\chNoiseTd(t)$ is additive white Gaussian noise (AWGN).
The received TD waveform is mapped back to the DD domain by the Zak transform,
$
\sigRxDD(\delay,\DopplerShift)
=
\mathcal{Z}_t\!\left(\sigRxTd(t)\right),
$
and subsequently filtered as
$
\sigRxDD^{\rxFilter}(\delay,\DopplerShift)
=
(\rxFilter\twistConv\sigRxDD)(\delay,\DopplerShift)
$
by the receive filter $w_\mathrm{rx}(\delay,\DopplerShift)$.
Sampling $\sigRxDD^{\rxFilter}$ on the lattice $\Lambda_{\dd}$ yields the discrete received signal
\begin{equation}\textstyle
\sigRxDD[\delayIdx,\DopplerIdx]
=
\sigRxDD^{\rxFilter}\!
\left(
    \delay = \frac{\delayIdx\delayPeriod}{\numDelayBin},\
    \DopplerShift = \frac{\DopplerIdx\DopplerPeriod}{\numDopplerBin}
\right),
\quad \delayIdx,\DopplerIdx \in \mathbb{Z}.
\label{eq:sampling}
\end{equation}
For the discrete received signals $\sigRxTd[i]$, the discrete Zak transform (DZT) is used to get $\sigRxDD[\delayIdx,\DopplerIdx]$:
\begin{equation}\label{eq:dzt}\textstyle
{\sigRxDDMatEle}[{\delayIdx},{\DopplerIdx}]
=
\frac{1}{\sqrt{\numDopplerBin}}
\sum_{i=0}^{{\numDopplerBin}-1}
    {\sigRxTd}[\delayIdx+i{\numDelayBin}]
    \cdot
    e^{-\jimg 2\pi \frac{{i\DopplerIdx}}{{\numDopplerBin}}}.
\end{equation}
We adopt the sinc shaping filters
$
\txFilter(\delay,\DopplerShift)
=
\rxFilter(\delay,\DopplerShift)
=
\sqrt{\bandwidth\frameTime}
\operatorname{sinc}(\bandwidth\delay)
\operatorname{sinc}(\frameTime\DopplerShift)
$~\cite{mohammed2023otfs}, corresponding to Nyquist sampling at a rate $\bandwidth$
over a frame of duration $\frameTime$ in discrete time,
so that the IDZT~\eqref{eq:idzt} and DZT~\eqref{eq:dzt} operate directly on the $\bandwidth\frameTime = \numDelayBin\numDopplerBin$ TD samples per frame without oversampling.

Combining \eqref{eq:lift}--\eqref{eq:sampling}, the transmit and receive pulse shaping can be incorporated into the effective DD channel
\begin{equation}
\chEff(\delay,\DopplerShift)
=
\rxFilter(\delay,\DopplerShift)
\twistConv
\chPhy(\delay,\DopplerShift)
\twistConv
\txFilter(\delay,\DopplerShift).
\end{equation}
Sampling the effective channel on $\Lambda_{\dd}$ gives the discrete effective channel
$
\chEff[\delayIdx,\DopplerIdx]
=
\chEff\!
\left(
    \delay = \delayIdx\frac{\delayPeriod}{\numDelayBin},\
    \DopplerShift = \DopplerIdx\frac{\DopplerPeriod}{\numDopplerBin}
\right).
$
The input--output relationship of the Zak-OTFS spread pilot system~\cite{ubadah2024zak}~is
\begin{equation}\label{eq:channel_io_heff}
\sigRxDDMatEle[\delayIdx,\DopplerIdx]
=
\chEffMatEle[\delayIdx,\DopplerIdx]
\twistConv
\sigTxDDMatEle[\delayIdx,\DopplerIdx]
+
\chNoiseDDMatEle[\delayIdx,\DopplerIdx],\
\forall \delayIdx, \DopplerIdx,
\end{equation}
where $\chNoiseDDMatEle[\delayIdx,\DopplerIdx]$ are the DD-domain noise samples.
Stacking $\numDelayBin\numDopplerBin$ received symbols $\sigRxDD[\delayIdx,\DopplerIdx]$, the matrix form of \eqref{eq:channel_io_heff} is
\begin{equation}\label{eq:channel_io_hdd}
{\sigRxDDVec} = {\chMatDD}{\sigTxDDVec} + {\chNoiseDDVec},
\end{equation}
where
${\sigTxDDVec} \in \mathbb{C}^{\numDelayBin\numDopplerBin\times1}$,
${\chNoiseDDVec} \in \mathbb{C}^{\numDelayBin\numDopplerBin\times1}$, and
$\chMatDD \in \mathbb{C}^{\numDelayBin\numDopplerBin\times\numDelayBin\numDopplerBin}$
denote the transmitted DD-domain signals, the noise vector, and the corresponding channel matrix, respectively~\cite{mohammed2023otfs}.
Note that $\sigTxDDMat=\text{vec}^{-1}(\sigTxDDVec) \in \mathbb{C}^{\numDelayBin\times\numDopplerBin}$ denotes the row-major DD-domain matrix.
The signal-to-noise ratio (SNR) is defined as the ratio of signal power (pilot plus data) to noise power.

\subsection{Spread Pilot Channel Sensing and Equalization}
\label{subsec:sys_xamb_eq}

For a single-pass receiver, the sensed effective channel $\chEffEstMat$ is extracted via a cross-ambiguity function~\cite{ubadah2024zak} of the DD-domain received signal $\sigRxDDMat$ and the known spread pilot $\sigTxDDSpMat$:
\begin{equation}\label{eq:heff_xamb}
\begin{aligned}
& \textstyle
    {\chEffEstMatEle}[\delayIdx,\DopplerIdx]
    =
    \xamb{\sigRxDDMat}{\sigTxDDSpMat}[\delayIdx,\DopplerIdx]
    =
    \frac{1}{\numDelayBin\numDopplerBin\sqrt{\pdr}} \\
& \textstyle
    \times
    \sum_{{\delayIdx}'=0}^{{\numDelayBin}-1}
    \sum_{{\DopplerIdx}'=0}^{{\numDopplerBin}-1}
        {\sigRxDDMatEle}[{\delayIdx}',{\DopplerIdx}']
        {\sigTxDDSpMatEle}^*[{\delayIdx}'-{\delayIdx},{\DopplerIdx}'-{\DopplerIdx}]
        e^{
            -\jimg2\pi
            \frac{{\DopplerIdx}(\delayIdx'-\delayIdx)}{{\numDelayBin\numDopplerBin}}
        }.
\end{aligned}
\end{equation}
The estimated channel matrix $\chMatDDEst$ is obtained via
\begin{equation}\label{eq:Hdd}
\begin{aligned}
& \textstyle
    {\chEleDDEst}[
        {\delayIdx}'{\numDopplerBin}+{\DopplerIdx}',
        {\delayIdx}{\numDopplerBin}+{\DopplerIdx}
    ]
    =
    \sum_{n,m\in\mathbb{Z}}
    e^{
        \jimg2\pi
        \frac{({\DopplerIdx}'-{\DopplerIdx}-m{\numDopplerBin})({\delayIdx}+n\numDelayBin)}{\numDelayBin {\numDopplerBin}}
    } \\
& \textstyle
    \times
    {\chEffEstMatEle}[
        {\delayIdx}'-{\delayIdx}-n\numDelayBin,
        {\DopplerIdx}'-{\DopplerIdx}-m{\numDopplerBin}
    ]
    e^{
            \frac{\jimg2\pi}{{\numDopplerBin}}n{\DopplerIdx}
    },
\end{aligned}
\end{equation}
where
${\delayIdx},{\delayIdx}'\in\delayIdxSet$
and
${\DopplerIdx},{\DopplerIdx}'\in\DopplerIdxSet$.

The data symbols are extracted from the superimposed frame by subtracting the pilot contribution, which is estimated from the known transmitted pilot and the sensed channel:
\begin{equation}\label{eq:pilot_cancel}
{\sigRxDDDtMatEst} = {\sigRxDDMat} - {\sigRxDDSpMatEst},
\
{\sigRxDDSpMatEst} \defn \text{vec}^{-1}(\sqrt{\pilotEnergy}{\chMatDDEst}{\sigTxDDSpVec}).
\end{equation}

The data equalization with linear minimum mean square error (LMMSE) uses the sensed channel matrix $\chMatDDEst$ from \eqref{eq:Hdd} and the vector form of estimated data symbols $\sigRxDDDtVecEst$ from \eqref{eq:pilot_cancel}:
\begin{equation}\label{eq:eq_lmmse}
{\sigTxDDDtVecEst}
=
\left[
    \left(
        {\chMatDDEstHerm}{\chMatDDEst}+{\noiseCovMat}
    \right)^{-1}
    {\chMatDDEstHerm}
\right]
{\sigRxDDDtVecEst}
\end{equation}
where ${\noiseCovMat} = ((1+\pdr)/\text{SNR}_\text{linear})\textbf{I}$ denotes the noise covariance matrix.
We solve~\eqref{eq:eq_lmmse} with the iterative conjugate gradient algorithm (CGA)~\cite{zheng2026real, mattu2025low} at complexity $\bigO(\cgaIters\numPath\numDelayBin\numDopplerBin)$ for $\numPath$ channel taps and $\cgaIters$ CGA iterations.

The equalized symbol is demodulated following
\begin{equation}\label{eq:demod}
\sigTxDDDtVecEstDemodEle
=
\mathcal{Q}_{\mathcal{S}}
    \left(
    {\sigTxDDDtVecEleEst}
    \right)
=
\arg\min_{s \in \mathcal{S}} \|{\sigTxDDDtVecEleEst} - s\|^2,
\end{equation}
$\forall\ {\sigTxDDDtVecEleEst} \in {\sigTxDDDtVecEst}$ and
$\mathcal{Q}_{\mathcal{S}}(\cdot)$ denotes the hard-decision QAM slicer over the constellation $\mathcal{S}$.
Demapping $\sigTxDDDtVecEstDemod$ into its corresponding binary sequence $\bitRx$ and comparing against the transmitted bits $\bitTx$ yields the BER.

\subsection{Iterative Pilot--Data Interference Cancellation}
\label{subsec:sys_turbo}

The single-pass receiver in Sec.~\ref{subsec:sys_xamb_eq} cancels the pilot contribution before equalization~\eqref{eq:pilot_cancel}, but leaves the data interference in channel sensing~\eqref{eq:heff_xamb}; turbo iteration~\cite{jayachandran2024zak} addresses this by canceling the data contribution estimated in the previous iteration before channel sensing.
At the $\turboIter$-th turbo iteration, the estimation of the received pilot signal $\atTurboIter{\sigRxDDSpMatEst}$ is
\begin{equation}\label{eq:turbo_data_cancel}
\atTurboIter{\sigRxDDSpMatEst}
=
\sigRxDDMat
-
\text{vec}^{-1}(
    \sqrt{\dataEnergy}
    \atLastTurboIter{\chMatDDEst}
    \atLastTurboIter{\sigTxDDDtVecEstDemod}
),
\end{equation}
where {\small$\atLastTurboIter{\sigTxDDDtVecEstDemod}$} is obtained by re-modulating the detected bits $\bitRx$.
Channel sensing~\eqref{eq:heff_xamb} is then repeated using $\atTurboIter{\sigRxDDSpMatEst}$ as
\begin{equation}\label{eq:turbo_heff_xamb}
\atTurboIter{\chEffEstMatEle}[\delayIdx,\DopplerIdx]
=
\xamb{\atTurboIter{\sigRxDDSpMatEst}}{\sigTxDDSpMat}[\delayIdx,\DopplerIdx].
\end{equation}
The resulting $\atTurboIter{\chMatDDEst}$~\eqref{eq:Hdd} from $\atTurboIter{\chEffEstMat}$ is used to cancel the pilot contribution~\eqref{eq:pilot_cancel}, followed by equalization~\cite{zheng2026real} and demodulation/re-modulation~\eqref{eq:demod} to obtain $\atTurboIter{\sigTxDDDtMatEstDemod}$.

%% file: 3_xamb.tex
\section{Low-Complexity Cross-Ambiguity}
\label{sec:xamb}

Direct evaluation of cross-ambiguity~\eqref{eq:heff_xamb} costs $\bigO(\numDelayBin^2\numDopplerBin^2)$.
This section derives two low-complexity algorithms by
\emph{(i)} exploiting the chirp structure of the OTFS spread pilot and 
\emph{(ii)} factorizing the twist matrix for arbitrary pilot structures.

\subsection{Chirp-Structured FFT-Based Cross-Ambiguity (CS-FFT)}\label{subsec:xamb_cs}

Inspired by the simplification of cross-ambiguity for a linear frequency modulation (LFM) chirp filter in Zak-OTFS-based radar application~\cite{mehrotra2025discrete}, we exploit the 2D quadratic chirp structure specific to the spread pilot to reduce \eqref{eq:heff_xamb} exactly to two Hadamard products, one 2D IFFT, and an index remapping.

\myparatight{Notation and assumptions.}
Let $\omega_L \defn e^{\jimg2\pi/L}$ and
$\varepsilon_L \defn 1$ for $L\equiv1\pmod4$ and
$\varepsilon_L \defn \jimg$ for $L\equiv3\pmod4$.
We use $\modinv{x}{L}$ to denote the multiplicative inverse of $x$
modulo $L$, and define the quadratic Gauss sum for an odd prime $L$ as
\begin{equation}\textstyle
\gaussSum(\alpha,L)
\defn
\sum_{r=0}^{L-1}\omega_L^{\alpha r^2}
=
\left(\frac{\alpha}{L}\right)\varepsilon_L\sqrt{L},
\end{equation}
with $\big(\frac{\alpha}{L}\big)$ the Legendre symbol.
We assume
\emph{(i)} $\numDelayBin$, $\numDopplerBin$ are distinct odd primes;
\emph{(ii)} $\gcd(\chirpSlope,\MN)=1$; and
\emph{(iii)} the point pilot has unit amplitude at the frame center
$(\delayIdxPilot,\DopplerIdxPilot)=\big(\frac{\numDelayBin-1}{2},\frac{\numDopplerBin-1}{2}\big)$.
The DD quasi-periodic extension follows \eqref{eq:qp_ext_disc}.

\myparatight{Closed-Form Spread Pilot.}
Within one $\numDelayBin\numDopplerBin\times\numDelayBin\numDopplerBin$ period, the quasi-periodic point pilot has $\numDopplerBin$ delay and $\numDelayBin$ Doppler replicas.
Substituting \eqref{eq:chirp_filter} into~\eqref{eq:spread_pilot} reduces the twisted convolution to two separable complete quadratic sums over the delay and Doppler wraps, which, by completing the square, satisfy
\begin{equation}\label{eq:quadratic_sum}\textstyle
\sum_{r=0}^{L-1}\omega_L^{\alpha r^2+\beta r}
=
\omega_L^{-\modinv{4\alpha}{L}\beta^2}
\gaussSum(\alpha,L).
\end{equation}
Consequently, the spread pilot has a 2D-chirp closed form
\begin{equation}\label{eq:xamb_cs_pilot}\textstyle
\sigTxDDSpMatEle[\delayIdx,\DopplerIdx]
=
\pilotConst\,\omegaMN^{\pilotPhase(\delayIdx,\DopplerIdx)},\quad
\pilotConst
\defn
\frac{
    \gaussSum(\chirpSlope\numDopplerBin,\numDelayBin)\,
    \gaussSum(\chirpSlope\numDelayBin,\numDopplerBin)
    }
{\sqrt{\MN}},
\end{equation}
$\forall \delayIdx,\DopplerIdx \in \mathbb{Z}$ so that $|\sigTxDDSpMatEle|\equiv|\pilotConst|=1$ (constant modulus),
and \(\pilotPhase\) is an integer quadratic polynomial:
\begin{equation}\label{eq:xamb_cs_phase}\small
\begin{aligned}
\pilotPhase(\delayIdx,\DopplerIdx)
&=\chirpSlope(\delayOff^2+\DopplerOff^2)+\delayIdxPilot\DopplerOff
  -\gM(2\chirpSlope\DopplerOff+\delayIdxPilot)^2
  -\gN(\DopplerIdx-2\chirpSlope\delayOff)^2\\
&=\pc{20}\delayIdx^2+\pc{02}\DopplerIdx^2+\pc{11}\delayIdx\DopplerIdx
  +\pc{10}\delayIdx+\pc{01}\DopplerIdx+\pc{00},
\end{aligned}
\end{equation}
where $\delayOff\defn\delayIdx-\delayIdxPilot$,
$\DopplerOff\defn\DopplerIdx-\DopplerIdxPilot$,
$\gM\defn\numDopplerBin\modinv{4\chirpSlope\numDopplerBin}{\numDelayBin}$,
$\gN\defn\numDelayBin\modinv{4\chirpSlope\numDelayBin}{\numDopplerBin}$,
and the subscripts of $\pc{ab}$ indicate the powers of $\delayIdx$ and $\DopplerIdx$, respectively.
Expanding~\eqref{eq:xamb_cs_phase} yields
\begin{equation}\label{eq:xamb_cs_coef}\small
\begin{aligned}
\pc{20} &= \chirpSlope-4\chirpSlope^2\gN,
\quad
\pc{02} = \chirpSlope-4\chirpSlope^2\gM-\gN,
\quad
\pc{11} = 4\chirpSlope\gN,\\
\pc{10} &= 2\chirpSlope\delayIdxPilot(\pc{11}-1),
\,
\pc{01} =
    (\delayIdxPilot-2\chirpSlope\DopplerIdxPilot)(1-4\chirpSlope\gM)
    -
    4\chirpSlope\delayIdxPilot\gN,\\
\pc{00} &=
    \chirpSlope(\delayIdxPilot^2+\DopplerIdxPilot^2)
    -\delayIdxPilot\DopplerIdxPilot
    -\gM(\delayIdxPilot-2\chirpSlope\DopplerIdxPilot)^2
    -4\chirpSlope^2\gN\delayIdxPilot^2.
\end{aligned}
\end{equation}
\noindent
From the definition of $\gM$ and $\gN$,
\begin{equation*}
\begin{alignedat}{2}
4\chirpSlope\gM &\equiv 1 \pmod{\numDelayBin},
&\qquad
\gM &\equiv 0 \pmod{\numDopplerBin},
\\
4\chirpSlope\gN &\equiv 1 \pmod{\numDopplerBin},
&\qquad
\gN &\equiv 0 \pmod{\numDelayBin}.
\end{alignedat}
\end{equation*}
Hence, the coefficients in~\eqref{eq:xamb_cs_coef} satisfy
\begin{equation}\label{eq:xamb_cs_div}
\begin{aligned}
\pc{20}&\equiv\pc{10}\equiv\pc{11}-1\equiv0
&&\pmod{\numDopplerBin},\\
\pc{02}&\equiv\pc{01}\equiv\pc{11}\equiv0
&&\pmod{\numDelayBin}.
\end{aligned}
\end{equation}

\myparatight{De-chirp and 2D-FFT Form.}
Substituting~\eqref{eq:xamb_cs_pilot} into \eqref{eq:heff_xamb}:
\begin{equation}\label{eq:xamb_cs_heff_1}\textstyle
\chEffEstMatEle[\delayIdx,\DopplerIdx]
=
\pilotConst^*
\frac{1}{\numDelayBin\numDopplerBin\sqrt{\pdr}}
\sum_{\delayIdx'=0}^{\numDelayBin-1}
\sum_{\DopplerIdx'=0}^{\numDopplerBin-1}
    \sigRxDDMatEle[\delayIdx',\DopplerIdx']
    \omegaMN^{\psi},
\end{equation}
with
$\psi
=
-\pilotPhase(\delayIdx'-\delayIdx,\DopplerIdx'-\DopplerIdx)
-\DopplerIdx(\delayIdx'-\delayIdx)$,
which can be expanded and collected into an input-dependent quadratic phase $\bar{\pilotPhase}$, terms linear in the summation indices
$(\delayIdx',\DopplerIdx')$ with frequencies $f_{\numDelayBin}$ and $f_{\numDopplerBin}$, and an output-dependent phase $\outPhase$:
\begin{equation}\label{eq:xamb_cs_expand}
\psi
=
-\bar{\pilotPhase}(\delayIdx',\DopplerIdx')
+\left(
    \delayIdx' f_{\numDelayBin}
    +
    \DopplerIdx' f_{\numDopplerBin}
\right)
-\outPhase(\delayIdx,\DopplerIdx),
\end{equation}
where $\bar{\pilotPhase}\defn\pilotPhase-\pc{00}$,
\begin{equation}\label{eq:xamb_cs_freq}
f_{\numDelayBin}
\defn
2\pc{20}\delayIdx+(\pc{11}-1)\DopplerIdx,
\ 
f_{\numDopplerBin}
\defn
\pc{11}\delayIdx+2\pc{02}\DopplerIdx,
\text{ and}
\end{equation}
\begin{equation}\label{eq:xamb_cs_outphase}
\outPhase(\delayIdx,\DopplerIdx)
\defn
\pc{20}\delayIdx^2+\pc{02}\DopplerIdx^2
+(\pc{11}-1)\delayIdx\DopplerIdx
-\pc{10}\delayIdx-\pc{01}\DopplerIdx+\pc{00}.
\end{equation}

By~\eqref{eq:xamb_cs_div} and~\eqref{eq:xamb_cs_freq},
$f_{\numDelayBin}/\numDopplerBin$ and
$f_{\numDopplerBin}/\numDelayBin$ are integers, reducing
\begin{equation}
\omegaMN^{\delayIdx'f_{\numDelayBin}}
=
e^{\jimg2\pi {\delayIdx}'(\frac{f_{\numDelayBin}}{{\numDopplerBin}})/{\numDelayBin}}
=
\omegaM^{\delayIdx'(\frac{f_{\numDelayBin}}{{\numDopplerBin}})},
\,
\omegaMN^{\DopplerIdx'f_{\numDopplerBin}}
=
\omegaN^{\DopplerIdx'(\frac{f_{\numDopplerBin}}{{\numDelayBin}})}
\end{equation}
to standard DFT kernels.
Since DFT frequency indices are periodic modulo their transform lengths, we define
\begin{equation}\label{eq:xamb_cs_idx}
\begin{aligned}
\idxM(\delayIdx,\DopplerIdx)
&\defn
\textstyle
\left\langle
\frac{f_{\numDelayBin}}{\numDopplerBin}
\right\rangle_{\numDelayBin}
=
\left\langle
\modinv{\numDopplerBin}{\numDelayBin}
(2\chirpSlope\delayIdx-\DopplerIdx)
\right\rangle_{\numDelayBin},\\
\idxN(\delayIdx,\DopplerIdx)
&\defn
\textstyle
\left\langle
\frac{f_{\numDopplerBin}}{\numDelayBin}
\right\rangle_{\numDopplerBin}
=
\left\langle
\modinv{\numDelayBin}{\numDopplerBin}
\left[
\delayIdx+
\left(
2\chirpSlope-\modinv{2\chirpSlope}{\numDopplerBin}
\right)\DopplerIdx
\right]
\right\rangle_{\numDopplerBin}.
\end{aligned}
\end{equation}
The corresponding kernels are therefore $\omegaM^{\delayIdx'\idxM}$ and $\omegaN^{\DopplerIdx'\idxN}$.

Define the de-chirped received signal as
\begin{equation}\label{eq:xamb_cs_dechirp}
\sigRxDechirp[\delayIdx',\DopplerIdx']
\defn
\sigRxDDMatEle[\delayIdx',\DopplerIdx']
\omegaMN^{-\bar{\pilotPhase}(\delayIdx',\DopplerIdx')}.
\end{equation}
Substituting~\eqref{eq:xamb_cs_expand} into \eqref{eq:xamb_cs_heff_1} then gives the cross-ambiguity
\begin{equation}\label{eq:xamb_cs_xamb_fft}\textstyle
\chEffEstMatEle[\delayIdx,\DopplerIdx]
=
\frac{\pilotConst^*}{\numDelayBin\numDopplerBin\sqrt{\pdr}}
\omegaMN^{-\outPhase(\delayIdx,\DopplerIdx)}
\sum\limits_{\delayIdx'=0}^{\numDelayBin-1}
\sum\limits_{\DopplerIdx'=0}^{\numDopplerBin-1}
    \sigRxDechirp[\delayIdx',\DopplerIdx']
    \omegaM^{\delayIdx'\idxM}
    \omegaN^{\DopplerIdx'\idxN},
\end{equation}
where the double sum is the unnormalized 2D IDFT of $\sigRxDechirp$.
Define the normalized 2D IDFT, computed by a 2D IFFT, as
\begin{equation}\label{eq:xamb_cs_ifft}\textstyle
\ifft\left\{\sigRxDechirp[\delayIdx',\DopplerIdx']\right\}[\idxM,\idxN]
\defn
\frac{1}{\MN}
\sum\limits_{\delayIdx'=0}^{\numDelayBin-1}
\sum\limits_{\DopplerIdx'=0}^{\numDopplerBin-1}
    \sigRxDechirp[\delayIdx',\DopplerIdx']
    \omegaM^{\delayIdx'\idxM}
    \omegaN^{\DopplerIdx'\idxN}.
\end{equation}
Then, the cross-ambiguity in \eqref{eq:heff_xamb} reduces to
\begin{equation}\label{eq:xamb_cs_final}
\boxed{
\begin{aligned}
& \textstyle
\chEffEstMatEle[\delayIdx,\DopplerIdx]
=
\frac{\pilotConst^*}{\sqrt{\pdr}}
\omegaMN^{-\outPhase(\delayIdx,\DopplerIdx)}
\\[-1mm]
&\times
\ifft
\!\left\{
\sigRxDDMatEle[\delayIdx',\DopplerIdx']
\omegaMN^{-\bar{\pilotPhase}(\delayIdx',\DopplerIdx')}
\right\}
[\idxM(\delayIdx,\DopplerIdx),
 \idxN(\delayIdx,\DopplerIdx)].
\end{aligned}
}
\end{equation}

\myparatight{Complexity and Implementation.}
The de-chirp table $\omegaMN^{-\bar{\pilotPhase}}$, the output table $\pilotConst^*\omegaMN^{-\outPhase}/\sqrt{\pdr}$, and the index maps~\eqref{eq:xamb_cs_idx} depend only on $(\numDelayBin,\numDopplerBin,\chirpSlope,\delayIdxPilot,\DopplerIdxPilot)$, which can be precomputed once and reused for every frame and every turbo iteration; $\sigRxDDMat$ is updated as~\eqref{eq:turbo_data_cancel}.
Each evaluation then costs $\bigO(\MN\log\MN)$.

\subsection{SVD-Factorized FFT-Based Cross-Ambiguity (SVD-FFT)}\label{subsec:xamb_svd}
For arbitrary pilot structures, the twist factor in~\eqref{eq:heff_xamb} precludes direct evaluation by a single 2D IFFT but can be factorized via singular value decomposition (SVD) to enable an FFT-based computation.
Define $\delayIdxSp\defn\delayIdx'-\delayIdx$ and $\DopplerIdxSp\defn\DopplerIdx'-\DopplerIdx$.
By~\eqref{eq:qp_ext_disc}, the summand of~\eqref{eq:heff_xamb} is periodic in $(\delayIdx',\DopplerIdx')$, so the summation can be shifted to $\delayIdxSp\in\delayIdxSet$, $\DopplerIdxSp\in\DopplerIdxSet$, with the remaining twist matrix given by
\begin{equation}\label{eq:xamb_svd_twist}
\twistMatEle[\DopplerIdx,\delayIdxSp]
\defn
\omegaMN^{-\DopplerIdx\delayIdxSp},
\quad
\DopplerIdx\in\DopplerIdxSetCtr,\
\delayIdxSp\in\delayIdxSet,
\end{equation}
where $\DopplerIdxSetCtr\defn\{-\lfloor\numDopplerBin/2\rfloor,\ldots,\lceil\numDopplerBin/2\rceil-1\}$, and likewise $\delayIdxSetCtr$, are centered index sets.
As a Vandermonde matrix with distinct nodes $\omegaMN^{-\DopplerIdx}$, $\twistMat$ has full rank, and its compact SVD gives
\begin{equation}\label{eq:xamb_svd_svd}\textstyle
\twistMat
=
\twistFacLMat\twistFacRMat^\mathrm{H}
=
\sum_{\twistCompIdx=1}^{\rank}
\twistFacLVec{\twistCompIdx}\twistFacRVec{\twistCompIdx}^\mathrm{H},
\quad
\rank=\min(\numDelayBin,\numDopplerBin),
\end{equation}
where $\twistFacLMat$ contains the scaled left singular vectors, and $\twistFacRMat$ the right singular vectors.
With the SVD-weighted pilot
$\sigTxDDSpCalMatEle{\twistCompIdx}[\delayIdxSp,\DopplerIdxSp]
\defn
\sigTxDDSpMatEle[\delayIdxSp,\DopplerIdxSp]\,
\twistFacRVecEle{\twistCompIdx}[\delayIdxSp]$,
\eqref{eq:heff_xamb} becomes
\begin{equation}\label{eq:xamb_svd_fft}
\begin{aligned}
&
\textstyle
\chEffEstMatEle[\delayIdx,\DopplerIdx]
=
\frac{1}{\numDelayBin\numDopplerBin\sqrt{\pdr}}
\sum_{\twistCompIdx=1}^{\rank}
\twistFacLVecEle{\twistCompIdx}[\DopplerIdx]
\\
&
\textstyle
\times
\sum_{\delayIdxSp=0}^{\numDelayBin-1}
\sum_{\DopplerIdxSp=0}^{\numDopplerBin-1}
    \sigRxDDMatEle[\delayIdxSp+\delayIdx,\DopplerIdxSp+\DopplerIdx]\,
    \sigTxDDSpCalMatEle{\twistCompIdx}^*[\delayIdxSp,\DopplerIdxSp],
\end{aligned}
\end{equation}
for $\delayIdx\in\delayIdxSetCtr$ and $\DopplerIdx\in\DopplerIdxSetCtr$.
Each inner sum is a 2D cross-correlation computed via FFTs.
Since $\delayIdxSp+\delayIdx$ spans $2\numDelayBin-1$ delay indices, we quasi-periodically extend $\sigRxDDMatEle$ via~\eqref{eq:qp_ext_disc} and zero-pad the SVD-weighted pilots to $2\numDelayBin$ to avoid circular-wrap contamination; the Doppler dimension needs no padding due to $\numDopplerBin$-periodicity.
The SVD and pilot FFTs are precomputed, so each evaluation requires one
2D FFT and $\rank$ 2D IFFTs of size $2\numDelayBin\times\numDopplerBin$, with complexity
$\bigO(\rank\MN\log\MN)$.

%% file: 4_impl.tex
\section{Implementation}\label{sec:impl}

\myparatight{Hardware/Software Platforms.}
We implement the Zak-OTFS processing pipeline with superimposed spread pilots, cross-ambiguity channel sensing, and turbo iterations based on the open-source GPU Zak-OTFS framework~\cite{zheng2026real}.
We precompute configuration-dependent quantities for SVD-FFT and CS-FFT, while the reported latency includes all online processing.
We evaluate the end-to-end pipeline on an NVIDIA RTX 6000 Ada GPU using CUDA Graphs to reduce kernel-launch overhead.
We benchmark cross-ambiguity latency on the GPU and a dual-socket server with Intel Xeon Gold 6530 CPUs.

\myparatight{Cross-Ambiguity Baselines.}
We implement the direct $\bigO(\numDelayBin^2\numDopplerBin^2)$ cross-ambiguity~\eqref{eq:heff_xamb} on both CPU and GPU, evaluating both loop-based and parallel implementations.
On the CPU, the parallel version distributes the independent $\chEffEstMatEle[\delayIdx,\DopplerIdx]$ across threads using OpenMP, while computing the inner summation over $(\delayIdx',\DopplerIdx')$ directly.
On the GPU, the loop-based implementation iterates over $\delayIdx$ while vectorizing the remaining dimensions, whereas the parallel implementation computes $\chEffEstMatEle[\delayIdx,\DopplerIdx]$ for all $(\delayIdx,\DopplerIdx)$ simultaneously using tensor broadcasting and reductions.
Full GPU parallelism requires materializing an $\bigO(\numDelayBin^2\numDopplerBin^2)$ intermediate tensor.

%% file: 5_eval.tex
\section{Performance Evaluation}\label{sec:eval}

\myparatight{Experiment Setup.}
We use the ITU Vehicular-A (Veh-A) channel model~\cite{itur_m1225} with six paths and a maximum delay of {$2.51$}\thinspace{\usec}.
The $\pathIdx$-th path has a fractional delay and a fractional Doppler shift $\pathDoppler=\chMaxDopplerFreq\cos\theta_i$, where $\theta_i$ is selected uniformly from $[0, 2\pi)$, independently for each frame, with the first path static.
We set the maximum Doppler shift to $\chMaxDopplerFreq=1$\thinspace{kHz}.
$\numDelayBin$ is chosen from odd primes near powers of two~\cite{ubadah2024zak}.
We set $\numDopplerBin=31$ and $\DopplerPeriod=30$\thinspace{kHz} to align with a common OFDM numerology, yielding a frame duration of $\frameTime=1.03$\thinspace{\msec} that serves as the latency budget of \emph{real-time} processing.
The corresponding delay and Doppler resolutions are $\delayRes=1/(\numDelayBin\DopplerPeriod)$ (e.g., $32.6$\thinspace{ns} at $\numDelayBin=1021$) and $\DopplerRes=968$\thinspace{Hz}, respectively.
For each ($\numDelayBin,\numDopplerBin$), we select the smallest admissible chirp slope $\chirpSlope$ achieving a normalized mean square error (NMSE) below $-100$\thinspace{dB} in auto-ambiguity recovery relative to the ideal point-pilot response.
Unless otherwise specified, we adopt a PDR of $-6$\thinspace{dB}~\cite{sheikh2024bayesian}, an SNR of $20$\thinspace{dB}, and QPSK modulation; the CGA equalizer uses $\cgaIters=10$ iterations~\cite{zheng2026real}.

\begin{figure}[!t]
    \centering
    \includegraphics[width=0.99\columnwidth]{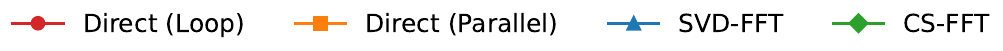}

    \vspace{-4.5mm}
    \subfloat[CPU]{
    \hspace{-3mm}
    \includegraphics[width=0.49\columnwidth]{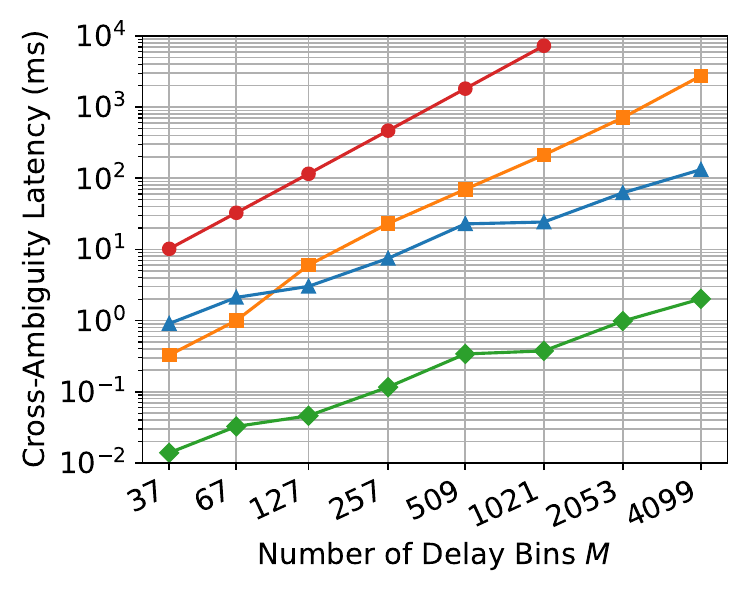}
    \label{subfig:eval_xamb_lat_cpu}}
    \subfloat[GPU]{
    \hspace{-3mm}
    \includegraphics[width=0.49\columnwidth]{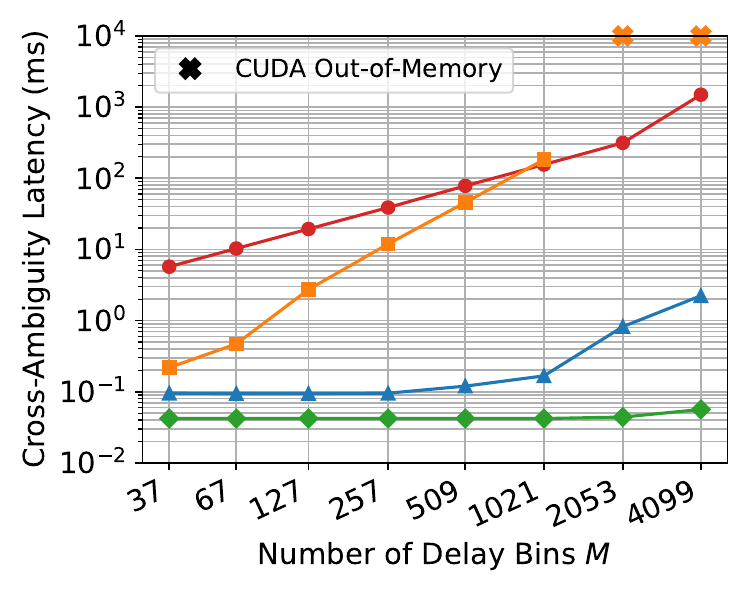}
    \label{subfig:eval_xamb_lat_gpu}}
    \caption{
    Latency of cross-ambiguity implementations ($\numDopplerBin=31)$.
    Direct (Parallel) uses CPU multithreading and GPU tensor vectorization.
    CS-FFT achieves the lowest latency and, on GPU, the best scalability among evaluated implementations.
    }
    \label{fig:eval_xamb_lat}
    \vspace{-1.5mm}
\end{figure}

\subsection{Cross-Ambiguity Latency}
Fig.~\ref{fig:eval_xamb_lat} compares the latency of cross-ambiguity implementations on CPU and GPU.
All implementations match the loop-based reference with NMSE below $5.97\times10^{-11}$.
We report median latency over $100$ trials after $10$ warm-up trials.
The direct method has complexity $\bigO(\numDelayBin^2\numDopplerBin^2)$ and is parallelized on both platforms.
On the CPU, 128-thread parallelism provides nearly constant speedup over the loop implementation while retaining similar latency scaling with $\numDelayBin$.
On the GPU, full vectorization incurs substantial memory overhead and eventually a CUDA out-of-memory (OOM) error.
Its latency scales less favorably with $\numDelayBin$ than the partial one with a loop.

Both the proposed CS-FFT and SVD-FFT scale well on the GPU.
Note that the time-consuming SVD~\eqref{eq:xamb_svd_svd} is computed offline.
From $\numDelayBin=67$ to $2053$ ($30.6\times$ bandwidth), CS-FFT latency grows only from $42.0$\thinspace{\usec} to $44.0$\thinspace{\usec}, reaching $56.3$\thinspace{\usec} at $\numDelayBin=4099$; in contrast, the loop-based implementation already requires $324.27$\thinspace{\msec} at $\numDelayBin=2053$.
For the spread-pilot use case, we use CS-FFT for the subsequent latency analysis.

\begin{figure}[!t]
    \centering
    \includegraphics[width=0.8\columnwidth]{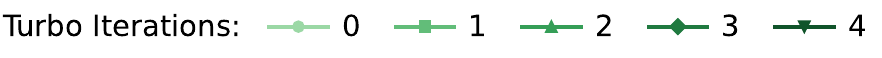}

    \vspace{-5mm}
    \subfloat[BER vs. M]{
    \hspace{-3mm}
    \includegraphics[width=0.49\columnwidth]{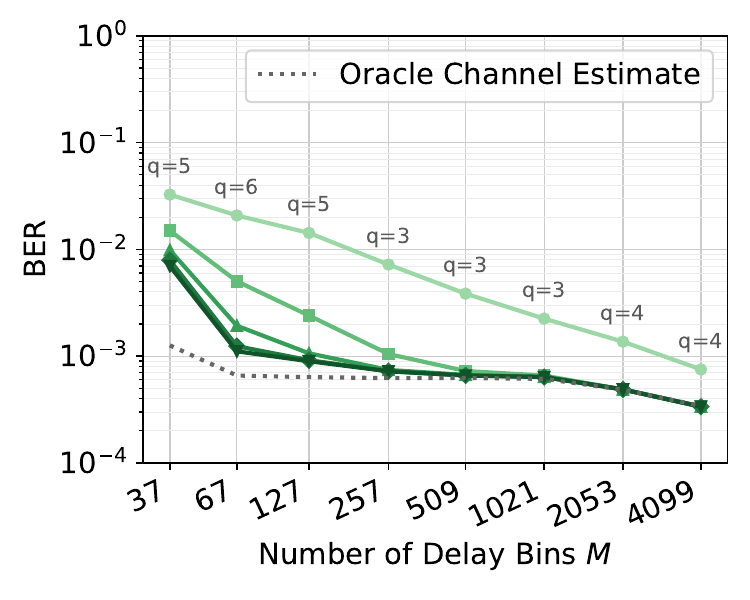}
    \label{subfig:e2e_ber}}
    \subfloat[Latency vs. M]{
    \hspace{-3mm}
    \includegraphics[width=0.49\columnwidth]{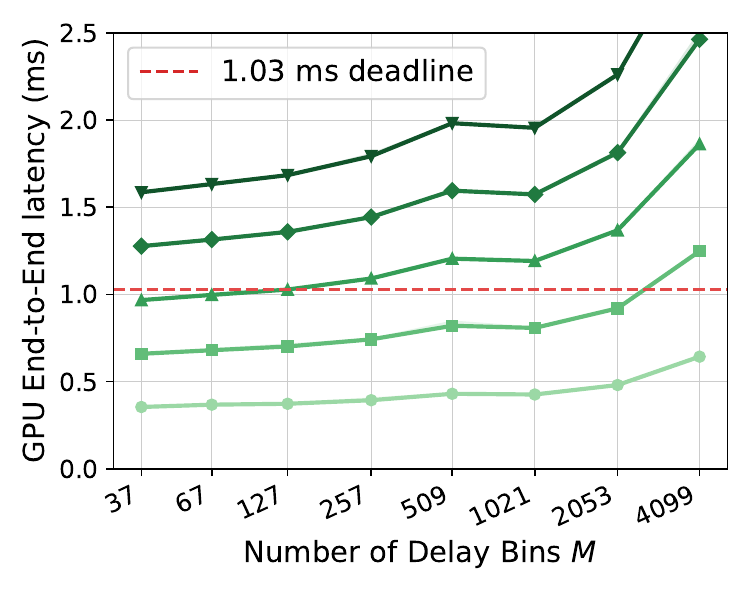}
    \label{subfig:e2e_lat}}
    \caption{
    End-to-end BER and latency over the number of delay bins $\numDelayBin$ with 20\,dB SNR and $-$6\,dB PDR.
    (a) BER gain diminishes after one iteration for $\numDelayBin\geq509$.
    (b) Each iteration adds constant latency for each $\numDelayBin$, while the pipeline meets the 1.03\thinspace{\msec} deadline for DD-domain grid dimension of up to $(\numDelayBin,\numDopplerBin) = (2053, 31)$ with only one iteration.
    }
    \label{fig:eval_e2e}
    \vspace{-1.5mm}
\end{figure}

\subsection{End-to-end System Performance}

Fig.~\ref{fig:eval_e2e} shows the BER and latency of the OTFS spread-pilot pipeline with turbo iterations versus the number of delay bins $\numDelayBin$.
The end-to-end pipeline includes discrete Zak transform, CS-FFT cross-ambiguity, $\chMatDDEst$ construction, pilot/data cancellation, equalization, and modulation/demodulation.

\myparatight{BER.}
Fig.~\ref{fig:eval_e2e}\subref{subfig:e2e_ber} evaluates the effect of turbo iterations across different frame sizes varying $\numDelayBin$, with oracle BER using a channel estimated from a noise-free pilot-only frame.
We set PDR $=-6$\thinspace{dB} here and examine its dependence on $\numDelayBin$ in Sec.~\ref{subsec:eval_pdr_turbo}.
Small frames keep benefiting from more iterations, whereas for $\numDelayBin\ge509$ the BER reaches the oracle after one iteration.
This is because the cross-ambiguity between the data signal and a spread pilot has magnitude on the order of $1/\sqrt{\pdr\numDelayBin\numDopplerBin}$~\cite{mehrotra2025zak}, so the data interference power on each estimated channel tap decreases as $1/(\pdr\numDelayBin\numDopplerBin)$.
For large frames, the first pilot/data cancellation therefore captures most of the gain, leaving little residual pilot--data interference for subsequent turbo iterations to remove; the same processing gain also permits a lower PDR (Sec.~\ref{subsec:eval_pdr_turbo}).
We hence use one turbo iteration as the operating point for the latency evaluation.

\myparatight{Latency.}
Fig.~\ref{fig:eval_e2e}\subref{subfig:e2e_lat} shows the GPU latency of the end-to-end baseband processing pipeline.
For a fixed $\numDelayBin$, latency increases approximately linearly with the number of turbo iterations, as each iteration repeats the receiver processing stages.
The real-time processing system meets the deadline of frame duration $1.03$\thinspace{\msec} up to $(\numDelayBin,\numDopplerBin)=(2053,31)$ with one turbo iteration.
With CS-FFT-accelerated cross-ambiguity, the pipeline latency is dominated by the sparsity-aware equalizer~\cite{zheng2026real}.

\begin{figure}[!t]
    \centering
    \includegraphics[width=0.8\columnwidth]{figs/lat_v_m_turbo_legend.pdf}

    \vspace{-5mm}
    \subfloat[$\numDelayBin=67$]{
    \hspace{-3mm}
    \includegraphics[width=0.49\columnwidth]{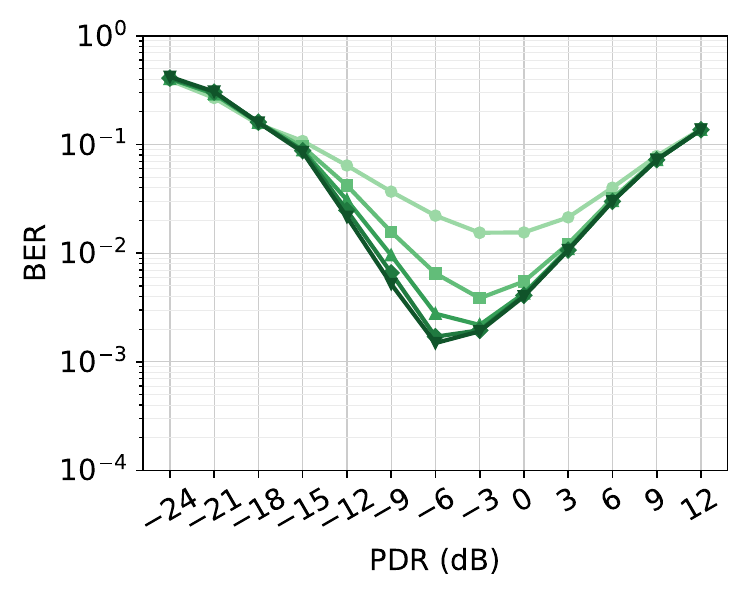}
    \label{subfig:ber_v_pdr_per_turbo_m67}}
    \subfloat[$\numDelayBin=2053$]{
    \hspace{-3mm}
    \includegraphics[width=0.49\columnwidth]{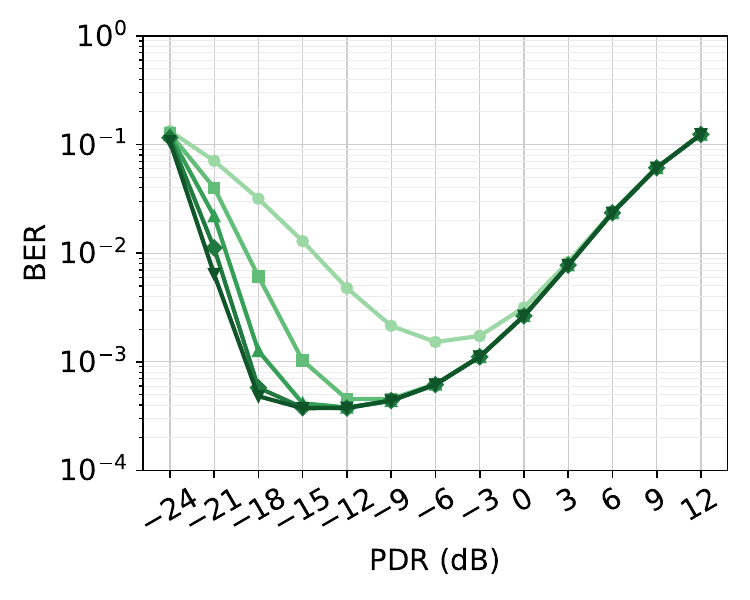}
    \label{subfig:ber_v_pdr_per_turbo_m2053}}
    \caption{
    BER vs. PDR at an SNR of $20$\thinspace{dB}.
    (a) The BER-minimizing PDR shifts toward lower values as the number of turbo iterations increases.
    (b) For $\numDelayBin=2053$, the overall BER-minimizing PDR shifts to about $-$15\thinspace{dB}, and curves with one or more iterations coincide for PDR $\geq-$9\thinspace{dB}.
    }
    \label{fig:ber_v_pdr_per_turbo}
    \vspace{-1.5mm}
\end{figure}

\subsection{Impact of PDR and Turbo Iterations}
\label{subsec:eval_pdr_turbo}

Higher pilot power improves the initial channel estimate, while lower pilot power leaves more power for data.
Turbo iterations improve BER but incur higher processing latency.
This section examines how the BER-optimal PDR and turbo convergence depend on the frame size and SNR.

\myparatight{Effect of Frame Size.}
Fig.~\ref{fig:ber_v_pdr_per_turbo} shows the BER versus PDR for $(\numDelayBin,\numDopplerBin)=(67,31)$ and $(2053,31)$ across turbo iterations at an SNR of $20$\thinspace{dB}.
For $\numDelayBin=67$ in Fig.~\ref{fig:ber_v_pdr_per_turbo}\subref{subfig:ber_v_pdr_per_turbo_m67}, with zero turbo iterations, the BER is minimized at a PDR around $-3$\thinspace{dB} and $0$\thinspace{dB}.
As turbo iterations are introduced, the BER-minimizing PDR shifts toward lower values, reaching $-6$\thinspace{dB} after three iterations, consistent with~\cite{sheikh2024bayesian}.
For $\numDelayBin=2053$ in Fig.~\ref{fig:ber_v_pdr_per_turbo}\subref{subfig:ber_v_pdr_per_turbo_m2053}, the BER-minimizing PDR shifts to $-9$\thinspace{dB} for one iteration and to $-15$\thinspace{dB} for more iterations.
For PDR $\geq-9$\thinspace{dB}, the BER converges after a single iteration, consistent with Fig.~\ref{fig:eval_e2e}\subref{subfig:e2e_ber}; at a lower PDR, additional iterations still improve BER but with diminishing returns.
At a PDR of $-9$\thinspace{dB}, one iteration reduces the BER by $4.67\times$,
from $2.13\times10^{-3}$ to $4.56\times10^{-4}$,
whereas four iterations at the best PDR of $-15$\thinspace{dB} reach $3.73\times10^{-4}$,
only 18\% lower despite three additional iterations.
Hence, under the real-time budget, a single iteration with a PDR around $-9$\thinspace{dB}
is near-optimal.
Therefore, the BER-optimal PDR depends on both frame size and the number of turbo iterations; no single PDR is optimal across configurations.

\begin{figure}[!t]
    \centering
    \includegraphics[width=0.99\columnwidth]{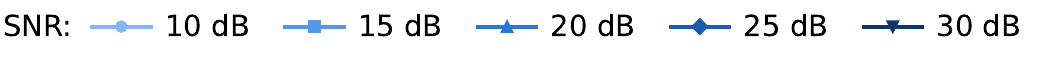}

    \vspace{-4.5mm}
    \subfloat[BER vs. PDR, 1 Turbo Iter.]{
    \hspace{-3mm}
    \includegraphics[width=0.49\columnwidth]{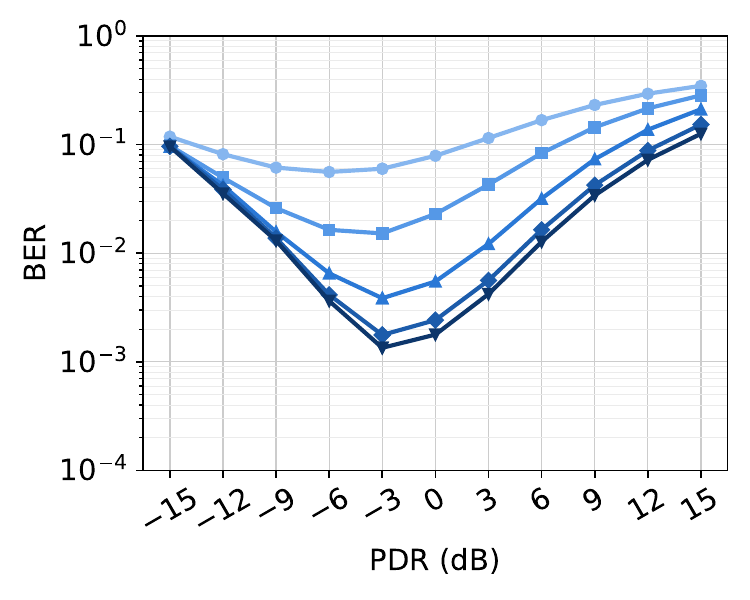}
    \label{subfig:ber_v_pdr_per_snr}}
    \subfloat[BER vs. Turbo Iter., PDR$=-6$\thinspace{dB}]{
    \hspace{-3mm}
    \includegraphics[width=0.49\columnwidth]{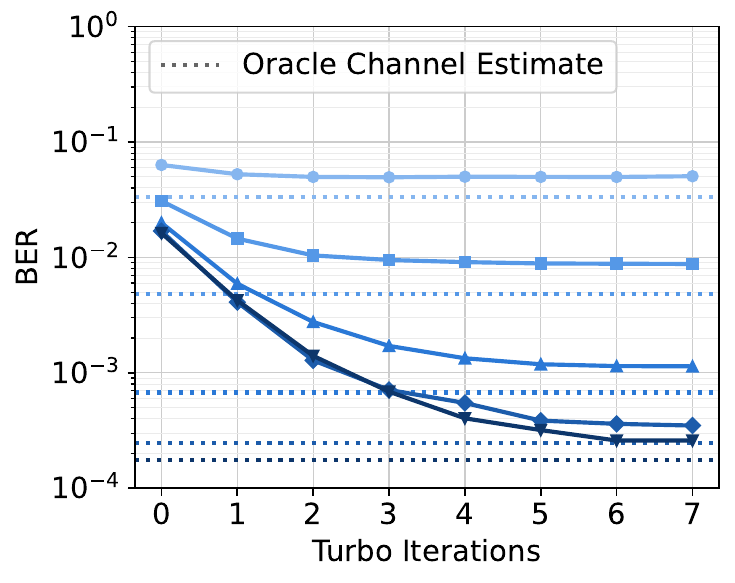}
    \label{subfig:ber_v_turbo_per_snr}}
    \caption{
    BER under multiple SNR for $(\numDelayBin,\numDopplerBin)=(67,31)$.
    (a) After one turbo iteration, the BER-minimizing PDR remains near $-3$\thinspace{dB} across SNRs.
    (b) At small $\numDelayBin$, the BER saturates after six turbo iterations, approaching the oracle-channel BER.
    }
    \label{fig:eval_ber_snr}
    \vspace{-1.5mm}
\end{figure}

\myparatight{Effect of SNR and Turbo Convergence.}
Fig.~\ref{fig:eval_ber_snr} focuses on a smaller frame, i.e., $(\numDelayBin,\numDopplerBin)=(67,31)$, where more turbo iterations continue to improve BER.
Fig.~\ref{fig:eval_ber_snr}\subref{subfig:ber_v_pdr_per_snr} sweeps PDR for multiple SNRs.
After one turbo iteration, the BER-minimizing PDR is less sensitive to SNR.
At moderate-to-high SNR (i.e., $15$\thinspace{dB} to $30$\thinspace{dB}), the system is interference-limited, and pilot--data interference rather than noise determines the optimal PDR at $-3$\thinspace{dB}; at $10$\thinspace{dB}, the curve flattens as noise becomes significant.
A similar trend is observed for other iterations, indicating that the BER-optimal PDR is relatively insensitive to SNR.
Fig.~\ref{fig:eval_ber_snr}\subref{subfig:ber_v_turbo_per_snr} shows turbo-iteration convergence under a PDR of $-6$\thinspace{dB}.
The BER continues to improve over several iterations before gradually saturating.
Together with Fig.~\ref{fig:ber_v_pdr_per_turbo}\subref{subfig:ber_v_pdr_per_turbo_m67}, the results show that increasing the PDR to $-3$\thinspace{dB} leads to faster convergence, with the BER
saturating within two iterations, but at a higher BER floor.
In contrast, the lower PDR of $-6$\thinspace{dB} allocates a larger fraction of the transmit power to data ($80\%$ vs.\ $67\%$), yielding a lower oracle BER and allowing additional turbo iterations to converge to a lower BER.

%% file: 6_concl.tex
\section{Conclusion}\label{sec:concl}

This paper presents a real-time GPU baseband processing pipeline for Zak-OTFS with superimposed spread pilots.
Exploiting the chirp structure of the spread pilot, we reduce the quasi-periodic cross-ambiguity exactly to a chirp-modulated 2D IDFT, lowering its complexity from $\bigO(\numDelayBin^2\numDopplerBin^2)$ to $\bigO(\numDelayBin\numDopplerBin\log\numDelayBin\numDopplerBin)$.
Our GPU implementation eliminates cross-ambiguity as the processing bottleneck and achieves end-to-end processing within a $1.03$\thinspace{\msec} frame time for $(\numDelayBin,\numDopplerBin)=(2053,31)$ with one turbo iteration.
We demonstrate that, as $\numDelayBin$ increases, the larger processing gain suppresses pilot--data interference, allowing one turbo iteration to capture most of the BER improvement at a lower PDR.